\documentclass[aip,graphicx]{revtex4-1}
\usepackage{float}
\makeatletter
\let\newfloat\newfloat@ltx
\makeatother
\usepackage{graphicx}% Include figure files
\usepackage{dcolumn}% Align table columns on decimal point
\usepackage{bm}% bold math
\usepackage[utf8]{inputenc}
\usepackage[T1]{fontenc}
\usepackage{mathptmx}
\usepackage{etoolbox}
\usepackage{algorithm}
\usepackage{algpseudocode}
\usepackage{amsmath}

\begin{document}

\title[]{FourPhonon\_GPU: A GPU-accelerated framework for calculating phonon scattering rates and thermal conductivity}

\author{Ziqi Guo}%
\affiliation{ 
School of Mechanical Engineering and the Birck Nanotechnology Center, Purdue University, West Lafayette, 47907, IN, USA%\\This line break forced with \textbackslash\textbackslash
}%

\author{Xiulin Ruan}%
 \email{ruan@purdue.edu}
\affiliation{ 
School of Mechanical Engineering and the Birck Nanotechnology Center, Purdue University, West Lafayette, 47907, IN, USA%\\This line break forced with \textbackslash\textbackslash
}%

\author{Guang Lin}%
 \email{guanglin@purdue.edu}
\affiliation{ 
School of Mathematics, Purdue University, West Lafayette, 47907, IN, USA%\\This line break forced with \textbackslash\textbackslash
}%


\date{\today}

\begin{abstract}

Accurately predicting phonon scattering is crucial for understanding thermal transport properties. However, the computational cost of such calculations, especially for four-phonon scattering, can often be more prohibitive when large number of phonon branches and scattering processes are involved. In this work, we present \texttt{FourPhonon\_GPU}, a GPU-accelerated framework for three-phonon and four-phonon scattering rate calculations based on the \texttt{FourPhonon} package. By leveraging OpenACC and adopting a heterogeneous CPU–GPU computing strategy, we efficiently offload massive, parallelizable tasks to the GPU while using the CPU for process enumeration and control-heavy operations. Our approach achieves over 25$\times$ acceleration for the scattering rate computation step and over 10$\times$ total runtime speedup without sacrificing accuracy. Benchmarking on various GPU architectures confirms the method's scalability and highlights the importance of aligning parallelization strategies with hardware capabilities. This work provides an efficient and accurate computational tool for phonon transport modeling and opens pathways for accelerated materials discovery.

\end{abstract}


\maketitle %\maketitle must follow title, authors, abstract and \pacs

\section{Introduction}

Thermal conductivity is a fundamental material property that plays a critical role in a wide range of applications, including thermal management in electronic devices~\cite{moore2014emerging,he2022experimental}, thermoelectric energy conversion~\cite{dresselhaus2007new,chen2003recent}, etc. Understanding and accurately predicting thermal conductivity is essential for optimizing material performance. The primary heat carriers in dielectrics and semiconductors are phonons. Phonon scattering rates, which limit the phonon mean free paths, are the central mechanism for determining thermal conductivity~\cite{ziman2001electrons}. Accurately predicting phonon scattering rates is therefore key to predicting thermal transport behavior in materials.

Recent advances in first-principles calculations of phonon scattering coupled with Boltzmann transport equation (BTE) have enabled accurate, parameter-free predictions of lattice thermal conductivity. The theoretical foundation for phonon BTE was first laid by Peierls~\cite{peierls1929kinetischen}. Later, Maradudin et al.\cite{maradudin1962scattering} developed a comprehensive framework for three-phonon (3ph) scattering. Broido et al.~\cite{broido2007intrinsic} further combined \textit{ab initio} force constants with these approaches, leading to robust first-principles predictions of thermal conductivity. It was believed for decades that 3ph scattering is adequate for describing thermal transport except at very high temperatures. Moreover, the general theory and computational method for four-phonon (4ph) scattering were lacking, preventing quantitative evaluation of its role. Feng and Ruan~\cite{feng2016quantum, feng2017four} developed the general theory and computational method for 4ph scattering, demonstrating that 4ph processes can significantly impact the thermal conductivity of many materials, at elevated temperatures or even room temperature. Their theoretical predictions for boron arsenide (BAs) were later confirmed experimentally~\cite{kang2018experimental, tian2018unusual, li2018high}, proving that four-phonon scattering is a critical factor in thermal transport. 

While first-principles calculations have greatly improved the fundamental understanding of thermal transport, they are extremely computationally expensive when including 4ph scattering.
The high cost arises from the need to enumerate and compute a large number of phonon scattering processes, which scale as $N^3$ and $N^4$ for 3ph and 4ph scattering, respectively ($N$ is the number of \textbf{q}-points in the Brillouin zone). This leads to a dramatic increase in computational complexity. For example, for a silicon calculation using a 16$\times$16$\times$16 \textbf{q}-mesh (discretized grid in the reciprocal space), there are approximately 9.0$\times 10^{5}$ and 7.6$\times 10^{9}$ processes for 3ph and 4ph scattering, respectively, which could take over 7000 CPU hours to calculate.

To address the high or even prohibitive computational cost, several approaches have been explored, including machine learning surrogates~\cite{guo2023fast, srivastava2024accelerating} and statistical sampling methods~\cite{guo2024sampling}. These techniques have been successfully applied to estimate thermal and radiative properties of solids~\cite{guo2024first, zhang2024cryogenic, guo2024firstoptical, tang2024effects, alkandari2025anisotropic, wei2024hierarchy, guo2025electronic, krutarth2025interface}. Although these methods dramatically reduce computational cost, they achieve this by introducing approximations that sacrifice some degree of accuracy.
As a result, they are particularly useful for applications where a certain level of error is acceptable. However, in scenarios that require rigorous and fully resolved calculations of every scattering process, such approximate methods are inadequate.

An alternative path for improving computational efficiency is leveraging hardware acceleration through Graphics Processing Units (GPUs). Originally developed for rendering computer graphics, GPUs have evolved as a powerful tool for scientific computing~\cite{owens2007survey}, machine learning~\cite{raina2009large,krizhevsky2012imagenet,carne2025overcomingcursedimensionalityenabling}, and high-performance simulations~\cite{stone2007accelerating} due to their massively parallel architecture. With thousands of cores capable of executing millions of lightweight threads simultaneously, GPUs are ideally suited for workloads that involve a large number of independent operations~\cite{kirk2016programming}. The advantage of GPUs has already been demonstrated across various fields of atomistic simulations. Popular first-principles simulation packages such as Abinit~\cite{gonze2016recent} and VASP~\cite{hacene2012accelerating, hutchinson2012vasp} have incorporated GPU support, achieving orders-of-magnitude acceleration over traditional CPU implementations. 
In the context of phonon scattering calculations, Wei et al.~\cite{wei2020towards} first identified the performance bottlenecks in \texttt{ShengBTE}~\cite{li2014shengbte} and offloaded the scattering rate calculations onto GPUs. Building on this foundation, Zhang et al.~\cite{zhang2021gpupbte} developed the \texttt{GPU\_PBTE} package, which employed a two-kernel strategy to further accelerate phonon scattering calculations.
It is shown that under the relaxation time approximation (RTA), each phonon scattering process is fully decoupled and can be independently evaluated, making the problem highly suitable for GPU parallelization.

\begin{figure}[h]%
\centering
\includegraphics[width=0.7\textwidth]{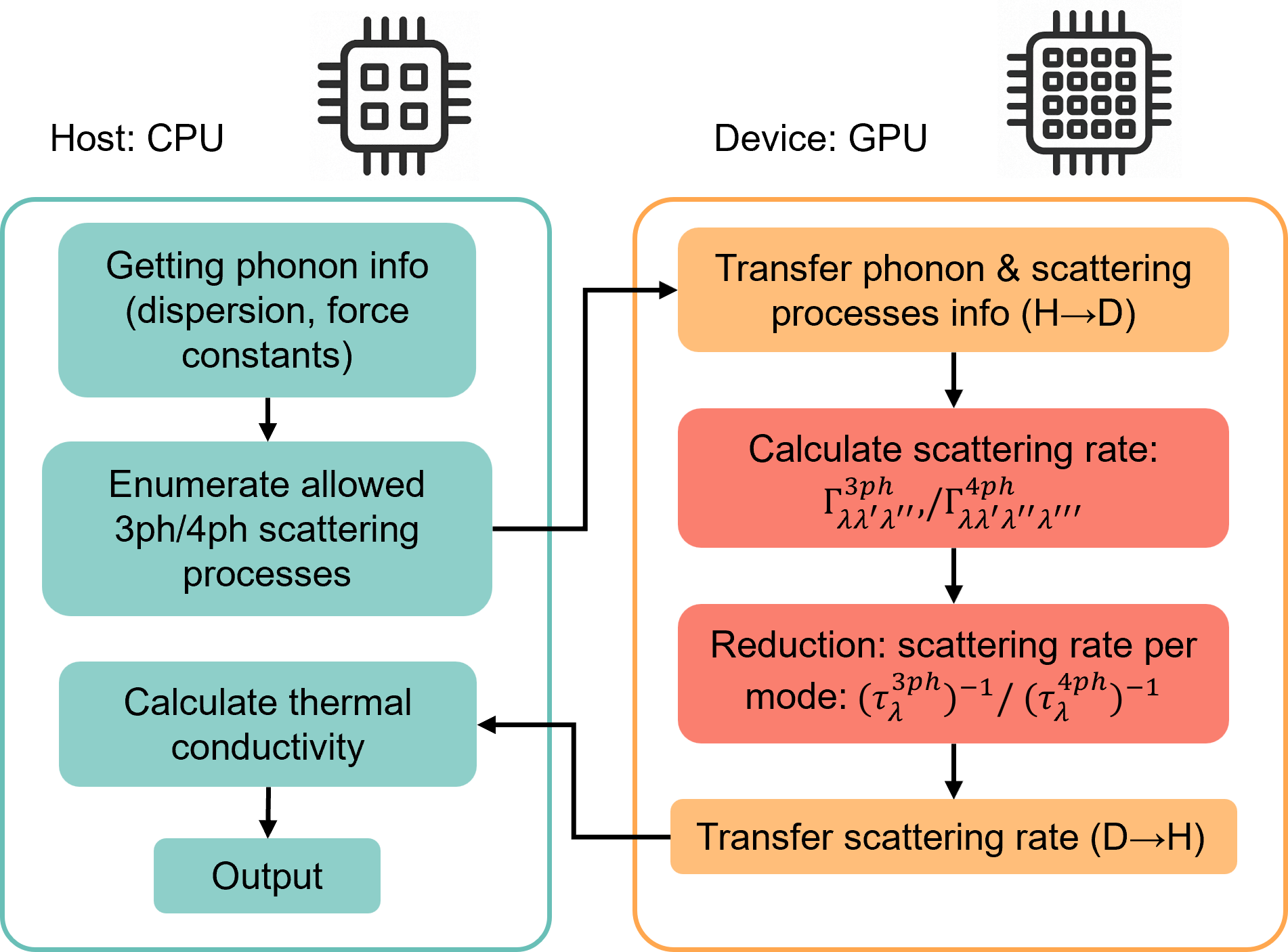} 
\caption{\textbf{Workflow of CPU-GPU heterogeneous computing.}}\label{Fig_flowchart}
\end{figure}

In this work, we develop a GPU-accelerated framework for both 3ph and 4ph phonon scattering calculations by combining CPU-based preprocessing with GPU-based large-scale parallel computing. Starting from the original \texttt{FourPhonon}~\cite{han2022fourphonon} package, we adopt OpenACC to enable GPU acceleration with minimal code restructuring. However, due to the challenges of divergent branching, direct GPU offloading alone is insufficient to achieve optimal performance. To overcome these limitations, we propose a heterogeneous CPU–GPU computing scheme, in which the CPU enumerates momentum- and energy-conserving scattering processes, and the GPU efficiently evaluates the corresponding scattering rates in parallel (Fig.~\ref{Fig_flowchart}). In addition, we implement several optimization techniques to enhance parallelism and further improve computational efficiency. 
Through systematic comparisons, we demonstrate that our method preserves the full accuracy of the original calculations while achieving over 25$\times$ acceleration for the scattering rate computation step and over 10$\times$ total speedup. Our work establishes an efficient and scalable pathway for rigorous phonon scattering calculations on modern high-performance computing architectures.
% among CPU-only, GPU-only, and heterogeneous implementations

\section{Methodology}

The original \texttt{ShengBTE}~\cite{li2014shengbte} and \texttt{FourPhonon}~\cite{han2022fourphonon} packages are written in Fortran and parallelized using MPI, optimized for CPU-based high-performance computing environments. To enable GPU acceleration with minimal code restructuring and maximize cross-platform portability, we adopt OpenACC, a directive-based programming model designed to simplify the development of heterogeneous applications targeting both CPUs and GPUs. Since the phonon scattering processes are independent of each other, there are no loop-carried dependencies and the task is highly suitable for GPU calculation. This computational pattern aligns well with the Single Instruction, Multiple Threads (SIMT) execution model of GPUs, where thousands of lightweight threads execute the same instruction on different data. As a result, the calculation of scattering rates across different scattering channels can be efficiently mapped onto parallel GPU threads.

To illustrate our GPU acceleration strategy, we use the 3ph absorption process as an example, with detailed pseudocode provided in the Appendix. In the original CPU implementation (Algorithm~\ref{algo-CPU_only} in the Appendix), the code loops over all phonon modes in a driver function, where each iteration calls a subroutine to evaluate the scattering rate for an individual mode. This subroutine performs nested loops over all possible combinations of phonon wavevectors and branches. Energy conservation checks are conducted to filter out forbidden scattering processes before computing the weighted phase space (WP) and scattering rate ($\Gamma$).
Due to the vast number of phonon combinations, this calculation is computationally expensive. Our first strategy is directly offloading the entire computation to the GPU (see Algorithm~\ref{algo-GPU_only} in the Appendix). All necessary data is preloaded into GPU memory to avoid host-device data transfer overhead during runtime. After GPU computation, the results are transferred back to CPU memory, and GPU memory is released. 
We parallelize over all possible scattering processes and, beyond that, across multiple phonon modes simultaneously to enhance concurrency. This all-modes parallelization approach outperforms the mode-by-mode parallelization used in prior GPU implementations~\cite{wei2020towards,zhang2021gpupbte} (see Algorithm~\ref{algo-GPU_hybrid-mode-wise}), offering higher levels of parallelism and improved speedup while having higher GPU memory cost. Both of these methods are implemented, and a comparison is provided in the Results section.

However, directly applying OpenACC directives to existing MPI-based Fortran code is insufficient due to architectural differences between CPUs and GPUs. Several optimizations have been implemented to make the code GPU-compatible and efficient. First, we apply loop flattening using the \texttt{collapse} clause to combine nested loops and expose more parallelism within each phonon mode. Second, the loop order is rearranged to achieve memory coalescing, allowing consecutive threads to access contiguous memory locations, which improves memory access efficiency. Third, we inline the computations for the broadening factor ($\sigma$) and the matrix element ($V_p$) to avoid the overhead of function and subroutine calls on the GPU. Finally, the accumulation of the scattering rate and weighted phase space is handled using the \texttt{reduction} clause, which efficiently manages parallel updates and avoids race conditions, outperforming the use of atomic operations.

\begin{figure}[h]%
\centering
\includegraphics[width=\textwidth]{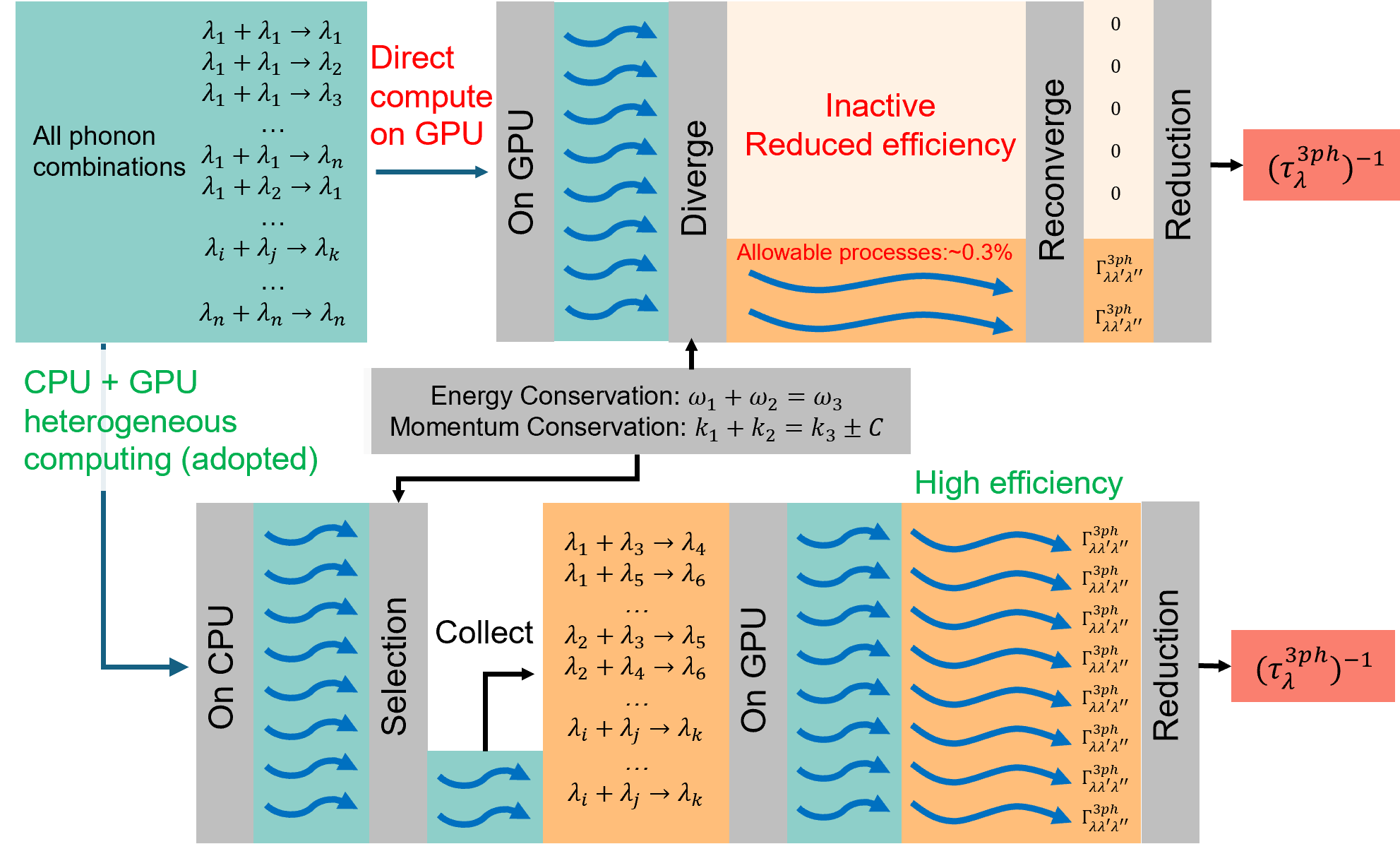} 
\caption{\textbf{GPU-only vs. CPU+GPU heterogeneous computing.} The GPU-only approach suffers from warp divergence due to conditional branching when filtering forbidden phonon scattering processes, leading to reduced computational efficiency. In contrast, the CPU+GPU heterogeneous strategy first filters and prepares valid scattering processes on the CPU, allowing the GPU to execute the computation with higher parallel efficiency.}\label{Fig_divergent_branching}
\end{figure}

\begin{figure}[h]%
\centering
\includegraphics[width=0.5\textwidth]{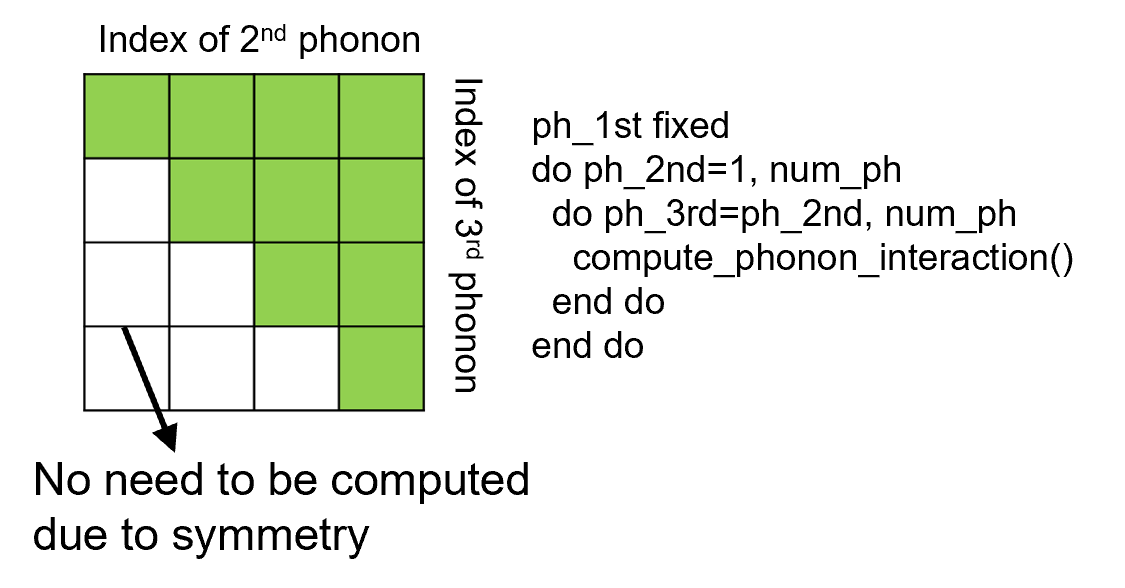} 
\caption{\textbf{Triangular iteration region due to symmetry.} }\label{Fig_symmetry}
\end{figure}

While this direct GPU-offloading approach preserves the original CPU workflow and achieves acceleration, it shows performance degradation due to divergent branching. Specifically, conditional statements used to check energy conservation introduce warp divergence, which undermines the efficiency of the SIMT execution. 
In our scenario, threads associated with forbidden scattering processes stall while others continue, leading to serialization and reduced throughput.
Given the sparsity of allowed scattering processes, this divergence leads to significant underutilization of GPU resources (Fig.~\ref{Fig_divergent_branching} upper figure). 
This problem has also been reported in prior work~\cite{zhang2021gpupbte}. Moreover, for 4ph scattering calculations, symmetry considerations restrict the iteration domain to a triangular region (see Fig.~\ref{Fig_symmetry}). However, OpenACC's \texttt{collapse} directive requires rectangular iteration domains, preventing the use of symmetry conditions on GPUs and further impacting performance.

To address these limitations, we further propose a hybrid CPU–GPU computing framework that partitions the workload between CPU and GPU (Fig.~\ref{Fig_divergent_branching} lower figure). In this scheme, the CPU is responsible for enumerating all symmetry-allowed and energy-conserving scattering processes, as described in Algorithm~\ref{algo-CPU_hybrid}. Once the allowable scattering process indices are generated, they are transferred to the GPU. The GPU then performs the expensive scattering rate calculations in parallel, following Algorithm~\ref{algo-GPU_hybrid}. By transferring only the relevant scattering events, we eliminate divergent branching on the GPU and ensure maximum occupancy of GPU resources during computation.
Although this hybrid approach introduces additional CPU overhead for process enumeration, it significantly reduces the computational cost associated with scattering rate evaluations and improves overall efficiency. In the Results section, we provide a detailed comparison between the CPU-only, GPU-only, and hybrid schemes, demonstrating the performance advantages of the proposed heterogeneous workflow. In our open source code, we adopted the hybrid approach for phonon scattering calculation.

\section{Results}

The performance of both the original and GPU-accelerated codes is evaluated using silicon as a benchmark material. All calculations are performed at 300 K under RTA for both 3ph and 3ph+4ph scattering processes. The broadening factor is chosen as 0.1 for 3ph+4ph calculation following our previous work~\cite{han2022fourphonon}. Simulations are carried out on the Gilbreth cluster at Purdue University's Rosen Center for Advanced Computing (RCAC). The software and hardware configurations are summarized in Table~\ref{tab:experimental-setup}.

\begin{table}[ht]
\centering
\caption{Experimental hardware and software configurations}
\begin{tabular}{ll}
\hline
Item & Description \\
\hline
CPU & AMD EPYC 7543 32-Core Processor \\
GPU & Primary: NVIDIA A100 (80GB) GPU. Tests: NVIDIA A10, A30 GPUs \\
Compiler (GPU) & NVIDIA HPC Compiler (nvc 23.5-0) \\
CUDA Version & CUDA 12.6.0 \\
Compiler (CPU) & Intel OneAPI Compilers 2024.2.1 \\
MPI Version & Intel MPI 2021.13 \\
\hline
\end{tabular}
\label{tab:experimental-setup}
\end{table}

For CPU-only calculations, parallel computing is performed with 32 CPU cores since the wall time for serial computing is impractical. The reported time is CPU time, which represents the summed computational time over all cores. For GPU-accelerated calculations, the simulations are performed using one CPU core and one GPU, and the reported runtime is the sum of the CPU and GPU computation times. The relative difference in the calculated thermal conductivity between CPU-only and GPU-accelerated runs is less than 0.1\%, which is primarily attributed to minor numerical variations between different compilers. These results confirm that GPU acceleration does not compromise the accuracy of the original code.

\begin{figure}[h]%
\centering
\includegraphics[width=1\textwidth]{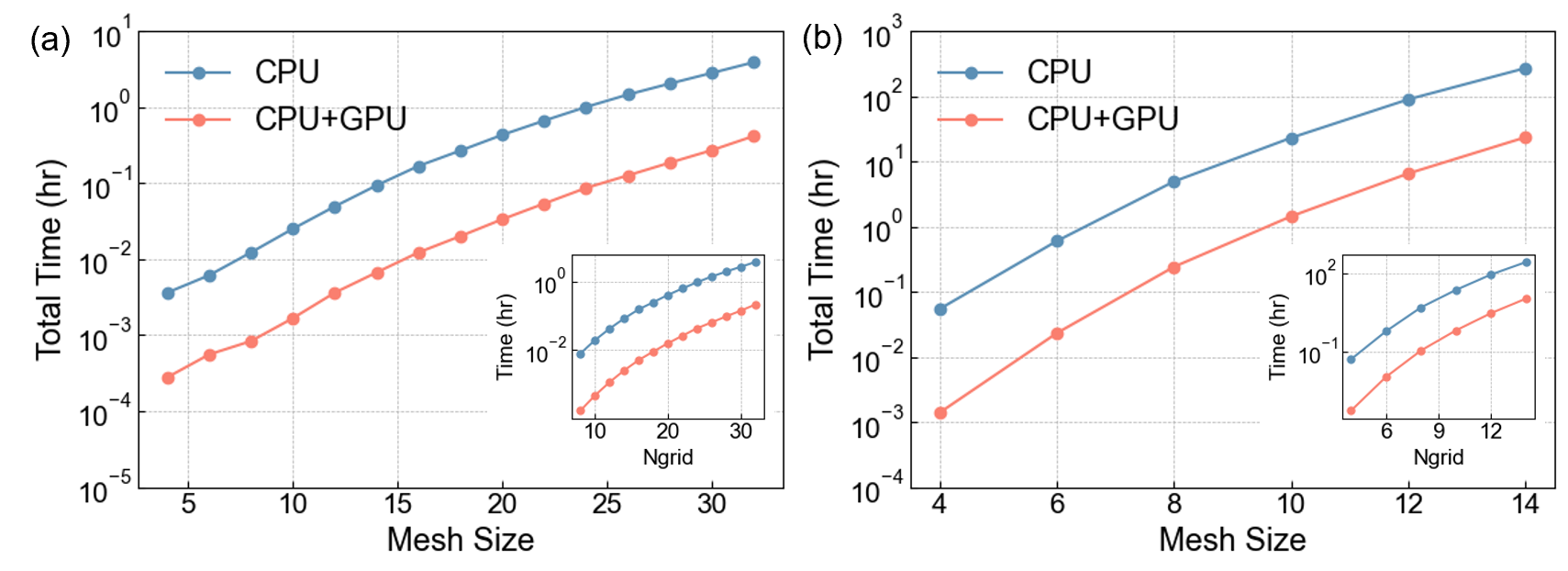} 
\caption{\textbf{Comparison of total computational cost between CPU-only and CPU-GPU hybrid implementations across different \textbf{q}-mesh sizes.} (a) 3ph scattering, (b) 3ph+4ph scattering. The insets show the isolated computational cost of the 3ph and 4ph scattering step alone. }\label{Fig_full_calc}
\end{figure}

We first analyze the acceleration achieved by our CPU-GPU heterogeneous computing implementation. Figure~\ref{Fig_full_calc} shows a comparison of the total computational cost between the original CPU-based method and the CPU-GPU hybrid version across different \textbf{q}-mesh densities.
For all tested \textbf{q}-meshes, we observe a consistent acceleration of over 10$\times$ for both 3ph and 3ph+4ph scattering calculations. 
Note that this computational time includes not only the phonon scattering step but also other computational overheads that are inherently performed on the CPU, such as the calculation of harmonic properties, phonon phase space, and the post-processing steps.
If we only consider the computational cost of the CPU+GPU phonon scattering calculation step, we observe even more substantial speedups of over 18$\times$ and over 25$\times$ for 3ph and 4ph scattering rate calculation, which is shown in the insets of Fig~\ref{Fig_full_calc}. 
We are expecting to see an even higher acceleration rate if we set the broadening factor to unity for the 3ph+4ph calculation. 
These results clearly demonstrate the effectiveness of our GPU acceleration strategy.

\begin{figure}[h]%
\centering
\includegraphics[width=\textwidth]{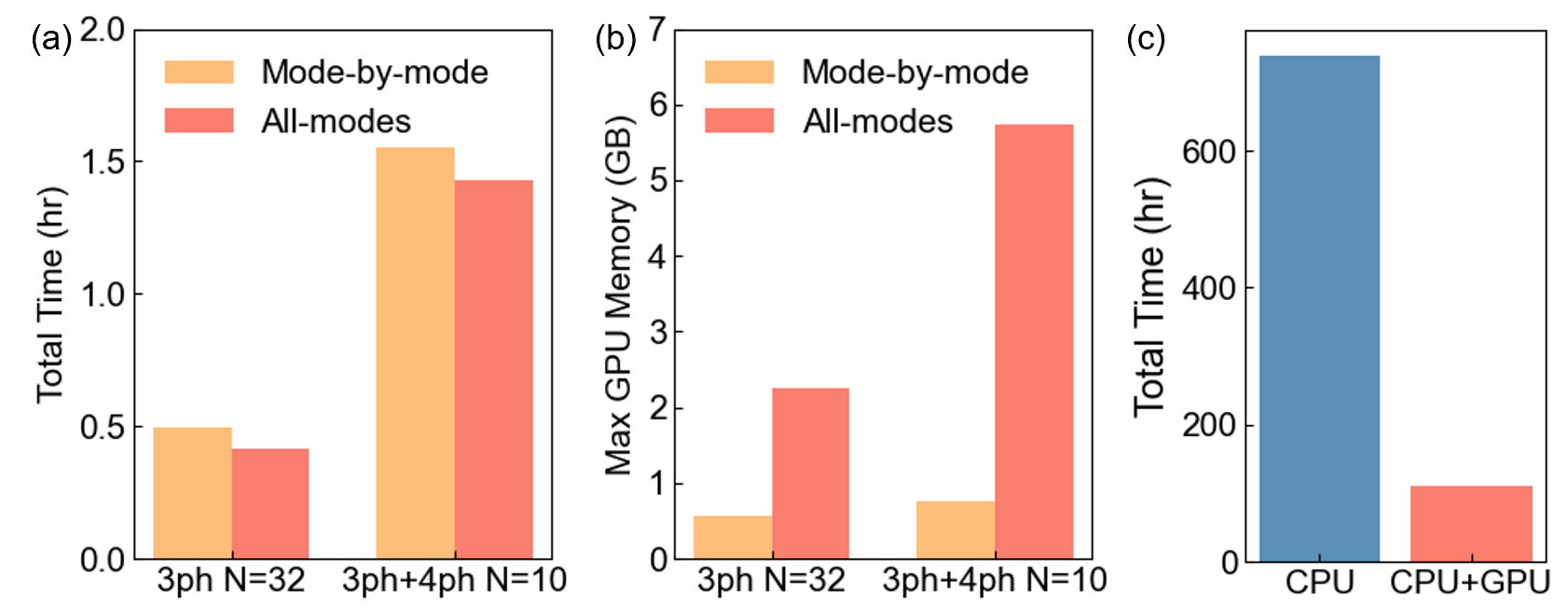} 
\caption{\textbf{Comparison between all-modes and mode-by-mode parallelization strategies.} (a) Computational cost and (b) GPU memory usage for 3ph and 3ph+4ph scattering calculations using a \textbf{q}-mesh of 32$\times$32$\times$32 and 10$\times$10$\times$10, respectively. (c) Computational cost for 3ph+4ph scattering with a 16$\times$16$\times$16 \textbf{q}-mesh, comparing CPU-only and CPU-GPU with mode-by-mode parallelization implementations.
}\label{Fig_mode_by_mode}
\end{figure}

We then compare two GPU parallelization strategies for phonon scattering calculations: mode-by-mode parallelization (where each phonon mode is processed independently) and all-modes parallelization (where all scattering processes are computed collectively). Figure~\ref{Fig_mode_by_mode}(a) shows the comparison for representative cases. Note that for 4ph scattering, we choose a smaller \textbf{q}-mesh size to illustradue to the large computational cost, but the trend would be similar for large mesh size. 
We observe that all-modes parallelization leads to additional acceleration of 21\% and 9\% for the 3ph scattering case with a 32$\times$32$\times$32 \textbf{q}-mesh and 3ph+4ph with 10$\times$10$\times$10 \textbf{q}-mesh, respectively. 
This performance gain is due to two factors: (1) enhanced parallelism enabled by concurrent execution of all scattering processes, and (2) reduced overhead in transferring data between CPU and GPU memory by avoiding frequent host–device memory traffic.
However, this gain comes with a trade-off. As shown in Fig.~\ref{Fig_mode_by_mode}(b), preloading all scattering processes into GPU memory significantly increases the GPU memory cost.  For dense \textbf{q}-meshes, this demand may exceed available GPU memory. This is the case for the 3ph+4ph calculation at a 16$\times$16$\times$16 mesh, which could not be completed using the all-modes strategy due to excessive memory usage (>80~GB).
To address this, we fall back to the mode-by-mode parallelization approach for the 16$\times$16$\times$16 case. While the computational efficiency is reduced, it still enables a notable acceleration of approximately 7$\times$ compared to the original CPU-based implementation (Fig.~\ref{Fig_mode_by_mode}(c)). This trade-off between memory usage and speed highlights the importance of selecting an appropriate parallelization strategy based on problem size and hardware constraints.

\begin{figure}[h]%
\centering
\includegraphics[width=0.7\textwidth]{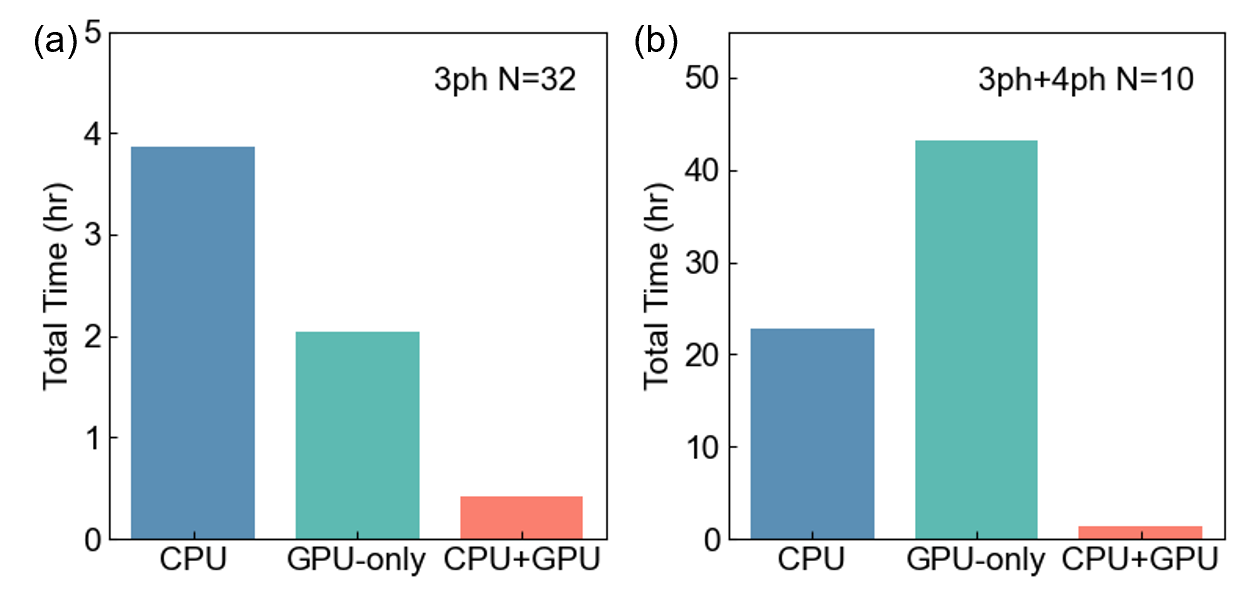} 
\caption{\textbf{Comparison between CPU–GPU hybrid and GPU-only implementations.} (a) 3ph scattering with \textbf{q}-mesh of 32$\times$32$\times$32, (b) 3ph+4ph scattering with \textbf{q}-mesh of 10$\times$10$\times$10. 
}\label{Fig_GPU_only}
\end{figure}

We further compared our CPU–GPU hybrid method with a GPU-only implementation, as illustrated in Fig.~\ref{Fig_GPU_only}.
For the 3ph scattering calculation with a 32$\times$32$\times$32 
\textbf{q}-mesh, while both approaches are faster than the CPU-only baseline, the CPU–GPU hybrid approach achieves a 5$\times$ speedup over the GPU-only method. Moreover, for the 3ph+4ph calculation using a \textbf{q}-mesh of 10$\times$10$\times$10, the GPU-only implementation is about twice as slow as the CPU-only baseline. 
This degradation is primarily due to the sparsity of the 4ph scattering matrix, which results in severe warp divergence and significantly reduces parallel efficiency.
These results further demonstrate the effectiveness of our heterogeneous computing strategy by using the GPU for highly parallel workloads while using CPU for irregular, branching-heavy operations.

\begin{figure}[h]%
\centering
\includegraphics{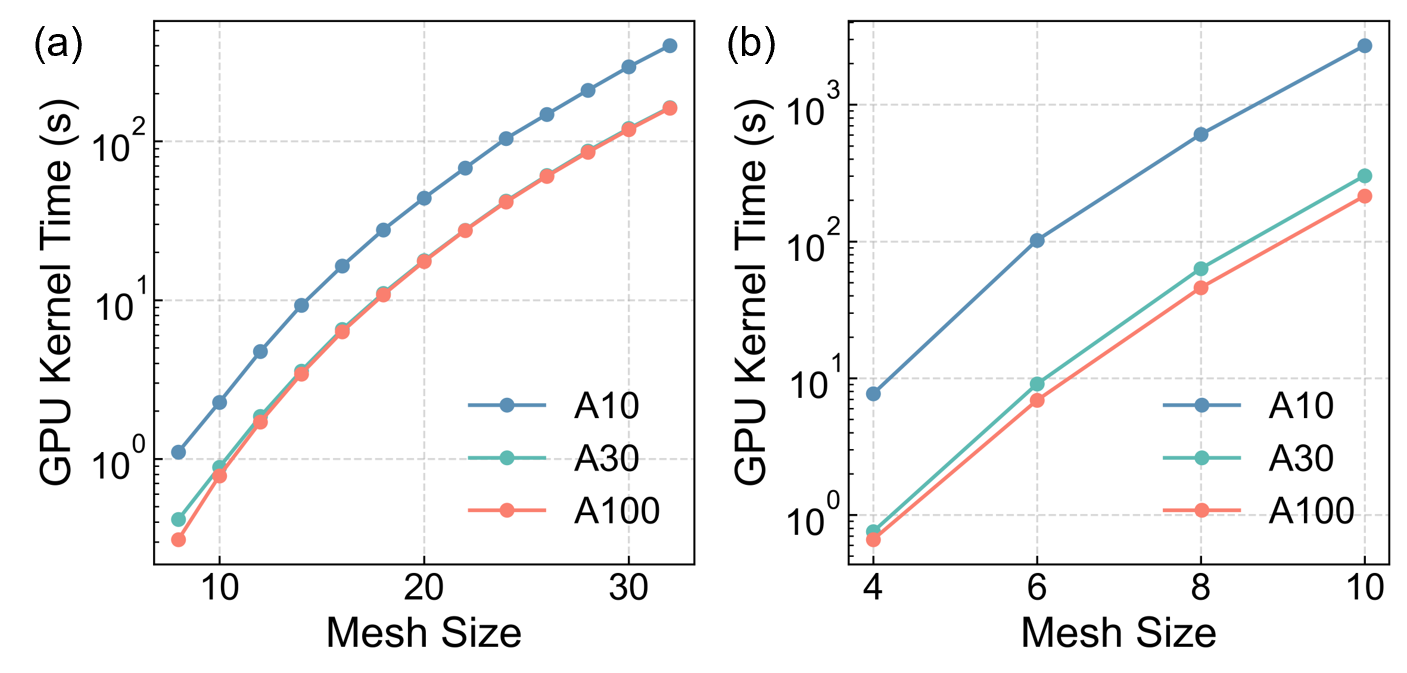} 
\caption{\textbf{Comparison of GPU kernel time on different NVIDIA GPUS (A10, A30, A100).} (a) 3ph scattering, (b) 4ph scattering. 
}\label{Fig_different_GPU}
\end{figure}

Finally, we evaluated the performance of our method on different GPU architectures. Since \texttt{FourPhonon} relies on double-precision arithmetic (FP64), the performance is strongly influenced by the available FP64 floating-point operations per second (FLOPS) on the GPU.
Specifically, we tested three NVIDIA GPUs: A100, A30, and A10. The key specifications of these GPUs, including memory capacity and peak FLOPS for both single-precision (FP32) and FP64 operations, are summarized in Table~\ref{tab:gpu_specs}. 
To isolate GPU performance, we measured only the execution time of the GPU kernel, excluding CPU-side overheads. 
We observe that A100 > A30 > A10 in terms of computational speed for our double-precision workloads (Fig.~\ref{Fig_different_GPU}), which is consistent with the FP64 FLOPS ranking.
Since the A10 is optimized for FP32 workloads and has significantly lower FP64 performance, it is less suitable for scientific computing applications and shows a substantial performance drop in our task.
Additionally, the A100’s larger memory capacity provides an advantage in dense \textbf{q}-mesh scenarios. For example, in the case of a 14$\times$14$\times$14 \textbf{q}-mesh with all-modes parallelization strategy, the total GPU memory requirement will be approximately 63 GB. This exceeds the memory capacity of the A10 and A30, which therefore will have to fall back to mode-by-mode parallelization, sacrificing performance to stay within hardware limitations.
These results highlight the importance of selecting GPU hardware that matches the computational precision and memory demands of the targeted simulation workload.

\begin{table}[h]
\centering
\caption{Specifications of selected NVIDIA GPUs used in this work.}
\label{tab:gpu_specs}
\begin{tabular}{lccc}
\hline
\textbf{GPU Model} & \textbf{FP32 Performance (TFLOPS)} & \textbf{FP64 Performance (TFLOPS)} & \textbf{Memory (GB)} \\
\hline
A100 & 19.5 & 9.7 & 80 \\
A30  & 10.3 & 5.2  & 24 \\
A10  & 31.2 & 0.39 & 24 \\
\hline
\end{tabular}
\end{table}

There are several directions for further improving the performance. Currently, our method is limited to the RTA due to the high memory cost and the overhead associated with data transfer between CPU and GPU memory. Future work could involve extending the framework to support iterative solvers. Besides, combining the GPU with the sampling-estimation-based approaches~\cite{guo2024sampling} would further reduce both time and memory cost. 
In addition, the enumeration step on the CPU and the parallel computing step on the GPU are executed sequentially. Introducing asynchronous operations to overlap the CPU and GPU calculations could further reduce total runtime. 
Lastly, while our implementation focuses on double-precision arithmetic for accuracy, it is worth noting that many modern GPUs, particularly those optimized for machine learning workloads, have significantly higher performance in single-precision computations. Exploring the use of mixed-precision or single-precision approaches, where appropriate, could yield additional performance gains without compromising accuracy for certain tasks.

In conclusion, we developed \texttt{FourPhonon\_GPU}, a GPU-accelerated framework for phonon scattering calculations using a CPU–GPU heterogeneous computing strategy. By offloading highly parallel tasks to the GPU and retaining enumeration and control-heavy operations on the CPU, our approach achieves substantial speedups, with over 10$\times$ improvement in total runtime and over 25$\times$ acceleration in the scattering rate calculation step. With comprehensive benchmarking tests, we demonstrated the effectiveness of this framework and highlighted the importance of aligning algorithm design with hardware capabilities. This work provides an efficient computational tool for evaluating materials' thermal properties and paves the way for accelerating materials discovery.

\clearpage

\section{Appendix}

\begin{algorithm}[H]
\caption{{\small Original CPU-based computation of three-phonon scattering process (absorption process)}}
{\footnotesize 
\begin{algorithmic}[1]
\State Initialize global arrays: rate\_scatt\_plus($N_{bands},N_{list}$), WP3\_plus\_array($N_{bands},N_{list}$)

\For{$mm = 1$ to $N_{bands}$*$N_{list}$}
    \State \texttt {// Subroutine: compute scattering for mode mm }

    \State Get $i, ll$ from $mm$
    \State Get $q$, $\omega$, $v$ from $i, ll$ %, skip if $\omega=0$
    \State Initialize $\text{WP3}_{\text{plus}}$$\gets 0$, $\Gamma_{\text{plus}}$$\gets 0$    
    \For{$j = 1$ to $N_{bands}$}
      \For{$ii = 1$ to $nptk$}
            \State Get $q'$, $\omega'$, $v'$ from $j, ii$ %, skip if $\omega'=0$        

        \State $f' \gets \text{BE}(\omega')$
    
        \For{$k = 1$ to $N_{bands}$}
            \State $q'' \gets \text{modulo}(q + q', N_{\text{grid}})$

          \State Get $ss$ from $q''$, get $\omega''$, $v''$ from $ss, k$  % , skip if $\omega''=0$

          \State $f'' \gets \text{BE}(\omega'')$
          \State $\sigma \gets \text{compute\_sigma}(v' - v'')$
    
          \If{$|\omega + \omega' - \omega''| \leq 2\sigma$}
            \State $\text{WP} \gets \frac{(f' - f'') \cdot \exp\left[-\frac{(\omega + \omega' - \omega'')^2}{\sigma^2}\right]}{\sigma \sqrt{\pi} \cdot \omega \omega' \omega''}$

            \State $\text{WP3}_{\text{plus}} \gets \text{WP3}_{\text{plus}} + \text{WP}$
    
              \State $V_p \gets \text{compute\_Vp}(\ldots)$
              \State $\Gamma_{\text{plus}} \gets \Gamma_{\text{plus}} + \text{WP} \cdot |V_p|^2$
          
          \EndIf
        \EndFor
      \EndFor
    \EndFor
    
    \State $\text{rate\_scatt\_plus}(i,ll) \gets \Gamma_{\text{plus}}$
    \State $\text{WP3\_plus\_array}(i,ll) \gets \text{WP3}_{\text{plus}}$
    \State \texttt {// End subroutine }

\EndFor
\State $\text{WP3\_plus\_array} \gets \text{WP3\_plus\_array} / nptk$
\State $\text{rate\_scatt\_plus} \gets \text{rate\_scatt\_plus} \cdot \text{const} / nptk$ \Comment{Unit conversion}

\end{algorithmic}\label{algo-CPU_only}
}
\end{algorithm}

\begin{algorithm}[H]
\caption{{\small GPU-only-based computation of three-phonon scattering process (absorption process)}}
{\footnotesize 
\begin{algorithmic}[1]

\State Initialize global arrays: rate\_scatt\_plus($N_{bands},N_{list}$), WP3\_plus\_array($N_{bands},N_{list}$)

\State {GPU: Copy data from host to device memory}
\State \texttt {// GPU: Launch parallel loop }
\For{$mm = 1$ to $N_{bands} \times N_{list}$}

    \State Get $i, ll$ from $mm$ \Comment{Inlined original subroutine for higher efficiency}
    \State Get $q$, $\omega$, $v$ from $i, ll$
    % \If{$\omega > \omega_{max}$} \textbf{continue} \EndIf
    \State Initialize $\text{WP3}_{\text{plus}}$$\gets 0$, $\Gamma_{\text{plus}}$$\gets 0$
    
    \State \texttt {// GPU: Launch parallel loop, collapse(3), reduction over $\Gamma_{\text{plus}}$, $\text{WP3}_{\text{plus}}$}
      \For{$ii = 1$ to $nptk$}
      \For{$j = 1$ to $N_{bands}$} \Comment{Reordered for memory coalescing}
        \For{$k = 1$ to $N_{bands}$} \Comment{Rearranged for GPU loop collapsing}
        
        \State Get $q'$, $\omega'$, $v'$ from $j, ii$

          \State $q'' \gets \text{modulo}(q + q', N_{grid})$
          \State Get $ss$ from $q''$, get $\omega''$, $v''$ from $ss, k$
            \State $f' \gets \text{BE}(\omega')$, $f'' \gets \text{BE}(\omega'')$
            
            \State \texttt {// GPU: parallel loop, reduction over $\sigma$}

          \State $\sigma \gets \text{inline\_compute\_sigma}(v' - v'')$ \Comment{Avoid function call for GPU efficiency}
          \If{$|\omega + \omega' - \omega''| \leq 2\sigma$}  \Comment{Divergent branching on GPU}
          
            \State $\text{WP} \gets \frac{(f' - f'') \cdot \exp\left[-\frac{(\omega + \omega' - \omega'')^2}{\sigma^2}\right]}{\sigma \sqrt{\pi} \cdot \omega \omega' \omega''}$
            \State $\text{WP3}_{\text{plus}} \gets \text{WP3}_{\text{plus}} + \text{WP}$
            
              \State $V_p \gets \text{inline\_compute\_Vp}(\ldots)$ \Comment{Avoid function call for GPU efficiency}
              \State $\Gamma_{\text{plus}} \gets \Gamma_{\text{plus}} + \text{WP} \cdot |V_p|^2$
          
          \EndIf
        \EndFor
      \EndFor
    \EndFor

    \State $\text{rate\_scatt\_plus}(i,ll) \gets \Gamma_{\text{plus}}$, $\text{WP3\_plus\_array}(i,ll) \gets \text{WP3}_{\text{plus}}$

\EndFor

\State $\text{WP3\_plus\_array} \gets \text{WP3\_plus\_array} / nptk$
\State $\text{rate\_scatt\_plus} \gets \text{rate\_scatt\_plus} \cdot \text{const} / nptk$ \Comment{Unit conversion}
\State {GPU: Copy data from device to host memory}

\end{algorithmic}\label{algo-GPU_only}
}
\end{algorithm}

\begin{algorithm}[H]
\caption{\small{CPU-based precomputing of three-phonon scattering process indices (absorption process)}}
{\footnotesize 
\begin{algorithmic}[1]
\State \texttt {// CPU: Enumerate of all allowed scattering processes}
\State Compute cumulative offset array $N_{\text{accum\_plus}}$ from $N_{\text{plus}}$

\State Initialize global arrays: $\text{Ind2}_{\text{all}}(\text{sum}(N_{\text{plus}}))$, $\text{Ind3}_{\text{all}}(\text{sum}(N_{\text{plus}}))$

\For{$mm = 1$ to $N_{bands} \times N_{list}$}
    \State Get $N_{plus}$ from $mm$
    
    \State Get $i, ll$ from $mm$
    \State Get $q$, $\omega$, $v$ from $i, ll$
    % \If{$\omega > \omega_{max}$} \textbf{continue} \EndIf
    \State Initialize $N_{\text{plus\_count}} \gets 0$
    \State Initialize arrays $\text{Ind2}(N_{\text{plus}})$, $\text{Ind3}(N_{\text{plus}})$

  \For{$ii = 1$ to $nptk$}
    \For{$j = 1$ to $N_{bands}$} \Comment{Reordered for memory coalescing}
        \State Get $q'$, $\omega'$, $v'$ from $j, ii$ %, skip if $\omega'=0$        
    
        \For{$k = 1$ to $N_{bands}$}
            \State $q'' \gets \text{modulo}(q + q', N_{\text{grid}})$

          \State Get $ss$ from $q''$, get $\omega''$, $v''$ from $ss, k$  % , skip if $\omega''=0$

          \State $\sigma \gets \text{compute\_sigma}(v' - v'')$
    
          \If{$|\omega + \omega' - \omega''| \leq 2\sigma$}
            \State $N_{\text{plus\_count}} \gets N_{\text{plus\_count}} + 1$
            \State $\text{Ind2}(N_{\text{plus\_count}}) \gets (ii - 1) \cdot N_{\text{bands}} + j$ \Comment{Detect possible scattering processes}
            \State $\text{Ind3}(N_{\text{plus\_count}}) \gets (ss - 1) \cdot N_{\text{bands}} + k$ \Comment{Detect possible scattering processes}
          \EndIf
        \EndFor
      \EndFor
    \EndFor
    \State Copy $\text{Ind2}$ to $\text{Ind2}_{\text{all}}$ at $N_{\text{accum\_plus}}(mm) + 1$
    \State Copy $\text{Ind3}$ to $\text{Ind3}_{\text{all}}$ at $N_{\text{accum\_plus}}(mm) + 1$

\EndFor

\end{algorithmic}\label{algo-CPU_hybrid}
}
\end{algorithm}

\begin{algorithm}[H]
\caption{{\small GPU-based computation of three-phonon scattering process using precomputed indices (absorption process)}}
{\footnotesize 
\begin{algorithmic}[1]

\State {GPU: Copy data from host to device memory}

% \State Build $Index\_N$ mapping from 3D $q$-grid to linear index
\State Compute cumulative offset array $N_{\text{accum\_plus}}$ from $N_{\text{plus}}$

\State \texttt{// GPU: Launch parallel loop}
\For{$mm = 1$ to $N_{\text{bands}} \times N_{\text{list}}$}

    \State Get $i, ll$ from $mm$
    \State Get $q$, $\omega$ from $i, ll$
  \State Initialize $\Gamma_{\text{plus}} \gets 0$, $\text{WP3}_{\text{plus}} \gets 0$

\State \texttt{// GPU: Launch parallel loop, reduction over $\Gamma_{\text{plus}}$, $\text{WP3}_{\text{plus}}$}
  \For{$ind$ from $N_{\text{accum\_plus}}(mm)+1$ to $N_{\text{accum\_plus}}(mm) + N_{\text{plus}}(mm)$}
  
    \State Get phonon indices: $(j,ii), (k,ss)$ from $\rm Ind2$, $\rm Ind3$
    \State Get $q'$, $\omega'$ from $j, ii$
    \State $q'' \gets \text{modulo}(q + q', N_{\text{grid}})$
    \State Get $\omega''$ from $k, ss$
    
    \State \texttt {// GPU: parallel loop, reduction over $\sigma$}

    \State Compute $\sigma \gets \text{inline\_compute\_sigma}(v' - v'')$ \Comment{Avoid function call for GPU efficiency}
    \State Compute $f' \gets \text{BE}(\omega')$, $f'' \gets \text{BE}(\omega'')$
    
    % \State \texttt{// Avoid divergent branching on GPU}
    
    \State $\text{WP} \gets \frac{(f' - f'') \cdot \exp\left[-\frac{(\omega + \omega' - \omega'')^2}{\sigma^2}\right]}{\sigma \sqrt{\pi} \cdot \omega \omega' \omega''}$
    \State $\text{WP3}_{\text{plus}} \gets \text{WP3}_{\text{plus}} + \text{WP}$

    \State $V_p \gets \text{inline\_compute\_Vp}(\ldots)$ \Comment{Avoid function call for GPU efficiency}
      \State $\Gamma_{\text{plus}} \gets \Gamma_{\text{plus}} + \text{WP} \cdot |V_p|^2$

  \EndFor

    \State $\text{rate\_scatt\_plus}(i,ll) \gets \Gamma_{\text{plus}}$
    \State $\text{WP3\_plus\_array}(i,ll) \gets \text{WP3}_{\text{plus}}$

\EndFor

\State $\text{WP3\_plus\_array} \gets \text{WP3\_plus\_array} / nptk$
\State $\text{rate\_scatt\_plus} \gets \text{rate\_scatt\_plus} \cdot \text{const} / nptk$ \Comment{Unit conversion}
\State {GPU: Copy data from device to host memory}

\end{algorithmic}\label{algo-GPU_hybrid}
}
\end{algorithm}

\begin{algorithm}[H]
\caption{{\small GPU-based, mode-wise computation of three-phonon scattering process using precomputed indices (absorption process)}}
{\footnotesize 
\begin{algorithmic}[1]

% \State Build $Index\_N$ mapping from 3D $q$-grid to linear index
\State Compute cumulative offset array $N_{\text{accum\_plus}}$ from $N_{\text{plus}}$

\For{$mm = 1$ to $N_{\text{bands}} \times N_{\text{list}}$}

    \State Get $i, ll$ from $mm$
    \State Get $q$, $\omega$ from $i, ll$
  \State Initialize $\Gamma_{\text{plus}} \gets 0$, $\text{WP3}_{\text{plus}} \gets 0$

\State {Copy mode-$mm$ data in $\rm Ind2$, $\rm Ind3$ from host to device memory} \Comment{Only copy mode-$mm$ for memory saving}

\State \texttt{// GPU: Launch parallel loop, reduction over $\Gamma_{\text{plus}}$, $\text{WP3}_{\text{plus}}$}
  \For{$ind$ from $N_{\text{accum\_plus}}(mm)+1$ to $N_{\text{accum\_plus}}(mm) + N_{\text{plus}}(mm)$}
  
    \State Get phonon indices: $(j,ii), (k,ss)$ from $Ind2$, $Ind3$
    \State Get $q'$, $\omega'$ from $j, ii$
    \State $q'' \gets \text{modulo}(q + q', N_{\text{grid}})$
    \State Get $\omega''$ from $k, ss$
    
    \State \texttt {// GPU: parallel loop, reduction over $\sigma$}

    \State Compute $\sigma \gets \text{inline\_compute\_sigma}(v' - v'')$ \Comment{Avoid function call for GPU efficiency}
    \State Compute $f' \gets \text{BE}(\omega')$, $f'' \gets \text{BE}(\omega'')$
    
    % \State \texttt{// Avoid divergent branching on GPU}
    
    \State $\text{WP} \gets \frac{(f' - f'') \cdot \exp\left[-\frac{(\omega + \omega' - \omega'')^2}{\sigma^2}\right]}{\sigma \sqrt{\pi} \cdot \omega \omega' \omega''}$
    \State $\text{WP3}_{\text{plus}} \gets \text{WP3}_{\text{plus}} + \text{WP}$

    \State $V_p \gets \text{inline\_compute\_Vp}(\ldots)$ \Comment{Avoid function call for GPU efficiency}
      \State $\Gamma_{\text{plus}} \gets \Gamma_{\text{plus}} + \text{WP} \cdot |V_p|^2$

  \EndFor
\State {Copy $\Gamma_{\text{plus}}$ and $\text{WP3}_{\text{plus}}$ of mode-$mm$ from device to host memory}

    \State $\text{rate\_scatt\_plus}(i,ll) \gets \Gamma_{\text{plus}}$
    \State $\text{WP3\_plus\_array}(i,ll) \gets \text{WP3}_{\text{plus}}$

\EndFor

\State $\text{WP3\_plus\_array} \gets \text{WP3\_plus\_array} / nptk$
\State $\text{rate\_scatt\_plus} \gets \text{rate\_scatt\_plus} \cdot \text{const} / nptk$ \Comment{Unit conversion}

\end{algorithmic}\label{algo-GPU_hybrid-mode-wise}
}
\end{algorithm}

\section*{Data availability}
The original results of the study are available from the corresponding authors upon reasonable request.

\section*{Code availability}
% The ShengBTE package integrated with the FourPhonon module is available at~\href{https://github.com/FourPhonon/FourPhonon}{https://github.com/FourPhonon/FourPhonon}. 

The work is incorporated as a new feature of the FourPhonon package and is available at~\href{https://github.com/FourPhonon/FourPhonon}{https://github.com/FourPhonon/FourPhonon}.

\section*{Competing interests}
The authors declare no competing interests.

\section*{Author contributions}
G.L., X.R. and Z.G. conceived the study. Z.G. designed and implemented the software, did the simulations, analyzed the results, and wrote the manuscript. G.L. and X.R. supervised the project. All authors contributed to discussions and revisions of the manuscript.

% \section*{Supplementary Material}

\begin{acknowledgments}
% Put your acknowledgments here.
Z.G. and X.R. acknowledge partial support from NSF Awards 2311848 and 2321301. Z.G is partly supported by the System Fellows 2024 Doctoral Fellowship provided by the Purdue Systems Collaboratory at Purdue University. Simulations were performed at the Rosen Center for Advanced Computing (RCAC) of Purdue University.
G.L. acknowledges the National Science Foundation under grants DMS-2053746, DMS-2134209, ECCS-2328241, CBET-2347401, and OAC-2311848. The U.S. Department of Energy also supports this work through the Office of Science Advanced Scientific Computing Research program (DE-SC0023161) and the Office of Fusion Energy Sciences (DE-SC0024583).
\end{acknowledgments}

\bibliography{sn-bibliography}% common bib file

\end{document}